\documentstyle[12pt,fleqn]{article}
\def\thebibliography#1{\section*{References}\list
  {[\arabic{enumi}]}{\settowidth\labelwidth{#1}\leftmargin\labelwidth
    \advance\leftmargin\labelsep
    \usecounter{enumi}}
    \def\newblock{\hskip .11em plus .33em minus .07em}
    \sloppy\clubpenalty4000\widowpenalty4000
    \sfcode`\.=1000\relax}

\def\op#1{\mathop{\fam0 #1}\limits}

\newcommand{\im}{{\rm Im\, }}
\newcommand{\nm}[1]{\mid {#1}\mid}

\newcommand{\beq}{\begin{equation}}
\newcommand{\eeq}{\end{equation}}
\newcommand{\ben}{\begin{eqnarray}}
\newcommand{\een}{\end{eqnarray}}
\newcommand{\be}{\begin{eqnarray*}}
\newcommand{\ee}{\end{eqnarray*}}
\newcommand{\bea}{\begin{eqalph}}
\newcommand{\eea}{\end{eqalph}}
\newcommand{\bR}{{\bf R}}
\newcommand{\al}{\alpha}
\newcommand{\bt}{\beta}
\newcommand{\la}{\lambda}
\newcommand{\f}{\phi}

\newcommand{\m}{\mu}

\newcommand{\g}{\gamma}
\newcommand{\G}{\Gamma}

\newcommand{\si}{\sigma}

\newcommand{\wt}{\widetilde}

\newcommand{\ol}{\overline}
\newcommand{\dr}{\partial}

\newcommand{\ar}{\op\longrightarrow}
\newcommand{\ot}{\otimes}

\newcounter{eqalph}
\newcounter{equationa}
\newcounter{theorem}
\newcounter{proposition}
\newcounter{lemma}
\newcounter{corollary}
\newcounter{definition}

\newenvironment{eqalph}{\stepcounter{equation}
\setcounter{equationa}{\value{equation}}
\setcounter{equation}{0}

\begin{eqnarray}}{\end{eqnarray}\setcounter{equation}{\value{equationa}}}

\def\thedefinition{\arabic{definition}}

\newenvironment{rem}{\medskip\noindent{\it Remark.}}{\medskip}

\newenvironment{prop}{\refstepcounter{definition} 
\bigskip\noindent{\it Proposition \thedefinition.}}{\medskip}
\newenvironment{lem}{\refstepcounter{definition} 
\bigskip\noindent{\it Lemma \thedefinition.}}{\medskip}
\newenvironment{cor}{\refstepcounter{definition} 
\bigskip\noindent{\it Corollary \thedefinition.}}{\medskip}

\begin{document}

\hbox{}

{\parindent=0pt 

{ \Large \bf Relativistic and non-relativistic equations of
motion}
\bigskip

{\sc Luigi Mangiarotti$\dagger$\footnote{E-mail
address: mangiaro@camserv.unicam.it} and Gennadi 
Sardanashvily$\ddagger$\footnote{E-mail address: sard@grav.phys.msu.su}}
\medskip

\begin{small}
$\dagger$ Department of Mathematics and Physics, University of Camerino, 62032
Camerino (MC), Italy \\
$\ddagger$ Department of Theoretical Physics, Physics Faculty, Moscow State
University, 117234 Moscow, Russia
\bigskip

{\bf Abstract.} It is shown that any second order dynamic equation on a
configuration space $X$ of non-relativistic time-dependent mechanics  can be
seen as a geodesic equation with respect to some (non-linear) connection on the
tangent bundle $TX\to X$ of relativistic velocities. Using this fact, the
relationship between relativistic and  non-relativistic equations of motion
is studied.
\end{small}
}

\section{Introduction}

In physical applications, one usually thinks of non-relativistic mechanics as
being an approximation of small velocities of a relativistic
theory. At the same time, the velocities in mathematical formalism of
non-relativistic mechanics are not bounded. It has long been
recognized that the relation between the mathematical schemes of
relativistic and non-relativistic mechanics is not trivial.  
Our goal is the following.

Let $X$ be a 4-dimensional world manifold of a relativistic theory,
coordinated by
$(x^\la)$. Then the tangent bundle $TX$ of $X$ plays the role of a space of
its 4-velocities (see Section 4).
A relativistic equation of motion is said
to be  a geodesic equation
\beq
\ddot x^\m=\dot x^\la\dr_\la\dot x^\m= K_\la^\m(x^\nu,\dot
x^\nu) \dot x^\la
\label{cqg2}
\eeq
with respect to a (non-linear) connection 
\beq
K=dx^\la\ot(\dr_\la +K^\m_\la\dot\dr_\m) \label{cqg3}
\eeq
on $TX\to X$. By $\dot x^\m(x)$ in (\ref{cqg2}) is meant a
geodesic vector field (which exists at least on a geodesic
curve), while $\dot x^\la\dr_\la$ is the formal operator of differentiation.
Throughout, we use the notation 
$\dr/\dr x^\la=\dr_\la$, $\dr/\dr \dot x^\la=\dot \dr_\la$. 

It is supposed additionally that there is a pseudo-Riemannian metric
$g$ of signature $(+,---)$ in $TX$ such that a geodesic vector field
does not leave the subbundle of
relativistic hyperboloids 
\beq
W_g=\{\dot x^\la\in TX\, \mid \,\,g_{\la\m} \dot x^\la\dot x^\m=1\}
\label{cqg1}
\eeq
in $TX$. It suffices to require that the condition
\beq
(\dr_\la g_{\m\nu}\dot x^\m + 2g_{\m\nu}K^\m_\la)\dot x^\la \dot x^\nu =0.
\label{cqg4}
\eeq
holds for all tangent vectors which belong to $W_g$ (\ref{cqg1}). 
 Obviously, the
Levi--Civita connection 
$\{_\la{}^\m{}_\nu\}$ of the metric $g$ fulfills the
condition (\ref{cqg4}). Any connection
$K$ on $TX\to X$ can be written as
\be
K^\m_\la = \{_\la{}^\m{}_\nu\}\dot x^\nu + \si^\m_\la(x^\la,\dot x^\la),
\ee
where the soldering form
\be
\si=\si^\m_\la dx^\la\ot\dot\dr_\la
\ee
plays the role of an external force. Then the condition (\ref{cqg4}) takes
the form
\beq
g_{\m\nu}\si^\m_\la\dot x^\la \dot x^\nu=0. \label{cqg46}
\eeq

Let now a world manifold $X$ admit a projection $X\to \bR$, where $\bR$ is
a time axis. One can think of the bundle $X\to\bR$ as being a configuration
space of a non-relativistic mechanical system. It is provided with the adapted
bundle coordinates $(x^0,x^i)$, where the transition functions of the temporal
one are $x'^0=x^0 +$const. The velocity phase space of a
non-relativistic mechanics is the first order jet manifold
$J^1X$ of $X\to \bR$, coordinated by $(x^\la, x^i_0)$
\cite{book,leon97a,book98,massa,sard98}. There is the canonical imbedding
of 
$J^1X$ onto the affine subbundle of the tangent
bundle $TX$, given by the coordinate condition 
\beq
\dot x^0=1, \qquad \dot x^i=x^i_0 \label{cqg7}
\eeq
(see (\ref{z260}) below).
Then one can think of (\ref{cqg7}) as the 4-velocities of a
non-relativistic system. The relation (\ref{cqg7}) differs from the relation
(\ref{cqg7'}) between 4- and 3-velocities of a relativistic system.
In particular, the
temporal component
$\dot x^0$ of 4-velocities of a non-relativistic system equals 1 (relative to
the universal unit system). It follows that the 4-velocities of
relativistic and non-relativistic systems occupy different subbundles of the
tangent bundle
$TX$. We show the following.

\begin{prop} \label{c1} Let $J^2X$ be the second order jet manifold
of $X\to\bR$, coordinated by $(x^\la,x^i_0,x^i_{00})$.  Each second order
dynamic equation 
\beq
x^i_{00} =\xi^i(x^0,x^j,x_0^j) \label{cqg5}
\eeq
of non-relativistic mechanics on $X\to\bR$ is equivalent to the geodesic
equation 
\ben
 && \dot x^\la\dr_\la\dot x^0 =0, \qquad \dot x^0=1, \nonumber\\
 && \dot x^\la\dr_\la\dot x^i= \ol K^i_0 \dot x^0 + \ol K^i_j\dot x^j
\label{cqg11}
\een
with respect to a connection
$\ol K$ on $TX\to X$ which fulfills the conditions
\beq
\ol K^0_\la =0, \qquad
 \xi^i= \ol K^i_0 + x^j_0\ol K^i_j\mid_{\dot x^0=1,\dot x^i=x^i_0}.
\label{cqg9}
\eeq 
\end{prop}

Note that, written relative to bundle coordinates
$(x^0,x^i)$ adapted to a given fibration $X\to\bR$, the connection $\ol K$
(\ref{cqg9}) and the geodesic
equation (\ref{cqg11}) are well defined with respect to any coordinates on
$X$. It should be also emphasized that the connection $\ol K$ (\ref{cqg9}) is
not determined uniquely. We will return repeatedly to this ambiguity. This
ambiguity is overcome if the relativistic transformation law of $\xi$ is
known.

Thus, we observe that both relativistic and non-relativistic equations of
motion can be seen as the geodesic equations on the same tangent bundle
$TX$.  The difference between them lies in the fact that their solutions
live in the different subbundles (\ref{cqg1}) and (\ref{cqg7}) of $TX$.  
At the same time, relativistic equations, expressed into the  3-velocities
$\dot x^i/\dot x^0$ of a relativistic system, tend exactly to the
non-relativistic  equations on the subbundle (\ref{cqg7}) when $\dot x^0\to
1$, $g_{00}\to 1$, i.e., where non-relativistic mechanics and the
non-relativistic approximation of a relativistic theory coincide only.

\section{Main relations}

There is the following relationship between relativistic and
non-relativistic equations of motion.

By a reference frame in non-relativistic
mechanics is meant an atlas of local constant trivializations of the bundle
$X\to\bR$ such that the transition functions of the spatial coordinates $x^i$
are independent of the temporal one $x^0$. In Section 3, we will give an
equivalent definition of a reference frame as a connection $\G$ on $X\to\bR$.

Given  a reference frame $(x^0,x^i)$, any connection $K(x^\la,\dot x^\la)$
(\ref{cqg3}) on the tangent bundle
$TX\to X$ defines the connection
$\ol K$ on $TX\to X$ with the components
\beq
\ol K^0_\la =0, \qquad \ol K^i_\la=K^i_\la. \label{cqg102}
\eeq
It follows that, given
a fibration $X\to\bR$, every relativistic equation of motion (\ref{cqg2})
yields the geodesic equation (\ref{cqg11}) and, consequently, has the
counterpart  
\beq
x^i_{00}=K^i_0(x^\la,1,x^k_0) +x^j_0K^i_j(x^\la,1,x^k_0) \label{cqg103}
\eeq
(\ref{cqg5}) in non-relativistic mechanics. 
Note that, written with respect to a
reference frame $(x^0,x^i)$, the connection $\ol K$ (\ref{cqg9}) and the
corresponding geodesic equation (\ref{cqg11}) are well defined relative to
any coordinates on
$X$, while the dynamic equation (\ref{cqg3}) is done relative to arbitrary
coordinates on $X$, compatible with the fibration $X\to\bR$. The key point is
that, for another reference frame $(x^0,x'^i)$ with
time-dependent transition functions $x^i\to x'^i$, the same connection $K$
(\ref{cqg3}) on $TX$ sets another connection $\ol K'$  on $TX\to X$ with the
components
\be
K'^0_\la=0, \qquad K'^i_\la = \left(\frac{\dr x'^i}{\dr x^j}K^j_\m +\frac{\dr
x'^i}{\dr x^\m}\right)\frac{\dr x^\m}{\dr x'^\la} +
\frac{\dr x'^i}{\dr x^0}K^0_\m \frac{\dr x^\m}{\dr x'^\la}
\ee 
with respect to the reference frame $(x^0,x'^i)$. It is easy to see that the
connection
$\ol K$ (\ref{cqg102}) has the components
\be
K'^0_\la=0, \qquad K'^i_\la = \left(\frac{\dr x'^i}{\dr x^j}K^j_\m +\frac{\dr
x'^i}{\dr x^\m}\right)\frac{\dr x^\m}{\dr x'^\la},
\ee
relative to same reference frame.
This illustrates the obvious fact that a non-relativistic approximation is not
relativistic invariant (see, e.g. \cite{levy}).

The converse procedure is more intricate. At first, a non-relativistic
dynamic equation (\ref{cqg5}) is brought into the geodesic equation
(\ref{cqg11}) with respect to the connection $\ol K$ (\ref{cqg9}). A solution
is not unique in general. Then, one should find a pair $(g,K)$  of a
pseudo-Riemannian metric
$g$ and a connection $K$ on $TX\to X$ such $K^i_\la=\ol K^i_\la$ and the
condition (\ref{cqg4}) is fulfilled. 

From the physical viewpoint, the most interesting dynamic equations are the
quadratic ones (sprays), i.e.,
\beq
\xi^i = a^i_{jk}(x^\m)x^j_0 x^k_0 + b^i_j(x^\m)x^j_0 + f^i(x^\m).
\label{cqg100}
\eeq
This property is global due to the transformation law (\ref{cqg26}).
Then one can use the following two facts.

\begin{prop}\label{c2} 
Any quadratic dynamic equation (\ref{cqg100}) is equivalent to the geodesic
equation (\ref{cqg11}) for the symmetric linear connection 
\be
\ol K=dx^\la\ot(\dr_\la + K_\la{}^\m{}_\nu(x^\al)\dot x^\nu\dot\dr_\m)
\ee
on $TX\to X$,
given by the components
\beq
K_\la{}^0{}_\nu=0, \quad K_0{}^i{}_0= f^i, \quad K_0{}^i{}_j=\frac12 b^i_j,
\quad K_j{}^i{}_k= a^i_{jk}. \label{cqg101}
\eeq
\end{prop}

It follows that every non-relativistic quadratic dynamic equation
\beq
x^i_{00}= a^i_{jk}(x^\m)x^j_0 x^k_0 + b^i_j(x^\m)x^j_0 + f^i(x^\m)
\eeq
(\ref{cqg100}) gives rise to the geodesic equation
\ben
&& \dot x^\la\dr_\la\dot x^0= 0, \qquad \dot x^0=1,\nonumber\\
&& \dot x^\la\dr_\la\dot x^i= 
a^i_{jk}(x^\m)\dot x^i \dot x^j + b^i_j(x^\m)\dot x^j\dot x^0 +
f^i(x^\m) \dot x^0\dot x^0 \label{cqg17}
\een
on a world manifold $X$.

\begin{prop}\label{c3}
Every affine vertical vector field 
\be
\si= (b^i_j(x^\m)x^j_0 + f^i(x^\m))\dr_i^0 
\ee
 on the affine jet bundle $J^1X\to X$ is extended to the soldering form
\be
\si=(f^idx^0 + b^i_kdx^k)\ot\dot \dr_i
\ee
on the tangent bundle $TX\to X$.
\end{prop}

It follows that, in particular, if there is no
topological obstruction and the Minkowski metric $\eta$ on $TX$ exists, a
non-relativistic dynamic equation 
\beq
x^i_{00}= b^i_j(x^\m)x^j_0 + f^i(x^\m)
\label{cqg18}
\eeq
gives rise to the geodesic equation
\ben
&& \dot x^\la\dr_\la\dot x^0= 0, \qquad \dot x^0=1, \nonumber\\
&& \dot x^\la\dr_\la\dot x^i=  b^i_j(x^\m)\dot x^j +
f^i(x^\m) \dot x^0. \label{cqg19}
\een

We meet the above-mentioned ambiguity.
The non-relativistic dynamic equation (\ref{cqg18}) can be
represented as both the geodesic equation (\ref{cqg19}) and the one
(\ref{cqg17}), where $a=0$. The first is the case for
external forces, e.g., an electromagnetic theory, while the latter is that for
a gravitation theory. One can also use the following assertion, instead of
Proposition \ref{c3}.

\begin{prop}
Any non-relativistic quadratic dynamic equation (\ref{cqg100}), being 
equivalent to the geodesic equation with respect to the linear connection
$\ol K$ (\ref{cqg101}), is also equivalent to the  one with respect an affine
connection $K'$ on $TX\to X$ which differs from $\ol K$ (\ref{cqg101}) in a
soldering form $\si$ on $TX\to X$ with the components
\be
\si^0_\la= 0, \qquad \si^i_k= h^i_k-\frac12 h^i_k\dot x^0, \qquad \si^i_0=
-\frac12 h^i_k\dot x^k -h^i_0\dot x^0 + h^i_0,
\ee
where $h^i_\la$ are local functions on $X$.
\end{prop} 

In Sections 3--4, we give a brief exposition of geometry of non-relativistic
and relativistic mechanics and prove Propositions
1--3 (see \cite{book,book98,sard98} for details). Section 5 is devoted to
several examples. In particular, we show that there is a coordinate system
where the Lagrange equation for a non-degenerate quadratic Lagrangian in
non-relativistic mechanics coincides with the non-relativistic approximation
of the geodesic motion in the presence of some pseudo-Riemannian or Riemannian
metric, whose spatial part is a mass tensor.

\section{Geometry of non-relativistic mechanics}

Let a fibre bundle $X\to\bR$, coordinated by $(x^0,x^i)$, be a configuration
space of non-relativistic mechanics. Its base $\bR$ is equipped with
the standard vector field $\dr_0$
and the standard 1-form
$dx^0$. The velocity phase space of non-relativistic mechanics is the
first order jet manifold $J^1X$ of sections $c$ of $X\to\bR$, which is
provided with the adapted coordinates $(x^0,x^i,x^i_0)$. Recall that $J^1X$
comprises the equivalence classes $j^1_{x^0}c$ of
$X\to\bR$ which are identified by their values  $c^i(x^0)$ and the values of
their derivatives
$\dr_0c^i(x^0)$ at points $x^0\in\bR$, i.e.,
\be
x^i_0(j^1_tc)= \dr_0c^i(t).
\ee 
There is the canonical imbedding 
\beq
\la: J^1X\hookrightarrow TX, \qquad \la=\dr_0 +x^i_0\dr_i,\label{z260}
\eeq
over $X$. 
From now on, we will 
identify $J^1X$ with its image in $TX$. 
It is an
affine bundle modelled over the vertical tangent bundle $VX$ of
$X\to\bR$. 

In
particular, every connection on a bundle $X\to\bR$ is given by the nowhere 
vanishing vector field 
\beq
\G:X\to J^1X\subset TX, \qquad \G=\dr_0 +\G^i\dr_i, \label{1005}
\eeq
on $X$. It can be treated as a reference frame in non-relativistic mechanics.
Every connection $\G^i$ (\ref{1005}) defines an atlas of local constant
trivializations of the bundle $X\to\bR$ and the associated coordinates
$(x^0,x^i)$ on $X$ such that the transition functions
$x^i\to x'^i$ are independent of
$x^0$,  and {\it vice versa} \cite{book}. We find
$\G^i=0$ with respect to these coordinates. In particular, there
is one-to-one correspondence between the complete connection
$\G$ (\ref{1005}) and the trivializations $X\cong
\bR\times M$ of the configuration
bundle $X$. Recall that different trivializations of $X$
differ from each other in projections of $X$ to its typical fibre $M$, while
the fibration $X\to\bR$ is one for all.

By a non-relativistic second order dynamic equation on a configuration bundle
$X\to\bR$ is meant the geodesic equation 
\be
x^i_{00}=\xi^i(x^\m,x^j_0) 
\ee
for a holonomic connection
\beq
\xi=\dr_0 + x^i_0\dr_i + \xi^i(x^\m,x^i_0) \dr_i^0 \label{a1.30}
\eeq
on the jet bundle $J^1X\to\bR$. This connection takes its values into
the second order jet manifold 
$J^2X\subset J^1J^1X$. It has the transformation law
\beq
\xi'^i=(\xi^j\dr_j + x^j_0x^k_0\dr_j\dr_k
+x^j_0\dr_j\dr_0 +\dr_0)x'^i.\label{cqg26}
\eeq

Let us consider the relationship
between the holonomic connections $\xi$ (\ref{a1.30}) on the jet bundle
$J^1X\to\bR$ and the connections 
\beq
 \g=dx^\la\ot (\dr_\la + \g^i_\la \dr_i^0)
\label{a1.38} 
\eeq 
on the affine jet bundle $J^1X\to X$. The connections $\g$ have the
transformation law
\beq
\g'^i_\la = (\dr_jx'^i\g^j_\m
+\dr_\m x'^i_0)\frac{\dr x^\m}{\dr x'^\la}. \label{m175}
\eeq

\begin{prop}\label{gena51} \cite{book,book98}.
Any connection $\g$ (\ref{a1.38}) on the affine jet bundle $J^1X\to X$ defines
the holonomic connection 
\beq
\xi = \dr_0 + x^i_0\dr_i +(\g^i_0 +x^j_0\g^i_j)\dr_i^0
\label{z281}
\eeq
on the jet bundle $J^1X\to\bR$.
\end{prop}

It follows that every connection $\g$ (\ref{a1.38}) on the affine jet bundle 
$J^1X\to X$ yields the dynamic equation
\beq
x^i_{00}=\g^i_0 +x^j_0\g^i_j \label{z287}
\eeq
on the configuration space $X$. 
Of course, different dynamic connections may lead to the same dynamic
equation (\ref{z287}).

\begin{prop}\label{gena52} \cite{book,book98}.
Any holonomic connection $\xi$ (\ref{a1.30}) on the jet bundle
$J^1X\to \bR$ defines a connection 
\beq
\g =dx^0\ot[\dr_0+(\xi^i-\frac12 x^j_t\dr_j^t\xi^i)\dr_i^0] +
dx^j\ot[\dr_j +\frac12\dr_j^0\xi^i \dr_i^0]
\label{z286}
\eeq
on the affine
jet bundle $J^1X\to X$.
\end{prop}

The connection $\g$ (\ref{z286}),
associated with a dynamic equation,  possesses the property
\beq
\g^k_i = \dr_i^0\g^k_0 +  x^j_0\dr_i^0\g^k_j
\label{a1.69}
\eeq
which implies $\dr_j^0\g^k_i = \dr_i^0\g^k_j$. 
Any connection $\g$, obeying the condition (\ref{a1.69}), is
said to be symmetric.

Let $\g$ be a connection (\ref{a1.38}) and $\xi_\g$ the 
corresponding dynamic equation (\ref{z281}). Then 
the connection (\ref{z286}), associated with $\xi_\g$, takes the form
\be
\g_{\xi_\g}{}^k_i = \frac{1}{2}
(\g^k_i + \dr_i^0\g^k_0 + x^j_0\dr_i^0\g^k_j),
\qquad \g_{\xi_\g}{}^k_0 = \xi^k - x^i_0\g_{\xi_\g}{}^k_i. 
\ee
It is readily observed that $\g = \g_{\xi_\g}$ if and only if $\g$ is
symmetric.

Since the jet bundle $J^1X\to X$ is affine, it admits an affine
connection 
\be
 \g=dx^\la\ot [\dr_\la + (\g^i_{\la 0}(x^\nu)+ \g^i_{\la
j}(x^\nu)x^j_0)\dr_i^0].
\ee
This connection is symmetric if and only if $\g^i_{\la \m}=\g^i_{\m\la}$. 
An affine connection $\g$ generates a quadratic dynamic equation, and {\it
vice versa}.
 
Now let us prove Proposition \ref{c1}.
We start from the relation between the connections $\g$ on the affine
jet bundle $J^1X\to X$ and the connections
$K$ (\ref{cqg3})
on the tangent bundle $TX\to X$ of the configuration space $X$.  Let us
consider the diagram
\beq
\begin{array}{rcccl}
& J^1_XJ^1X & \ar^{J^1\la} & J^1_XTX & \\
_\g &  \put(0,-10){\vector(0,1){20}} & &  \put(0,-10){\vector(0,1){20}}
& _K\\
& J^1X &\ar^\la & TX &
\end{array} \label{z291}
\eeq
where $J^1_XTX$ is the first order jet manifold of the tangent bundle $TX\to
X$, coordinated by $(x^\la,\dot x^\la,\dot x^\la_\m)$.
The jet prolongation over $X$ of the canonical imbedding $\la$
(\ref{z260}) reads 
\be
J^1\la: (x^\la,x^i_0, x^i_{\m 0}) \mapsto 
(x^\la,\dot x^0=1,\dot x^i=x^i_0, \dot x_\m^0=0,
\dot x^i_\m=x^i_{\m 0}).
\ee
We have
\be
&& J^1\la\circ \g: (x^\la,x^i_0) \mapsto 
(x^\la,\dot x^0=1,\dot x^i=x^i_0, \dot x_\m^0=0,
\dot x^i_\m=\g^i_\m ),\\
&& K\circ \la: (x^\la,x^i_0) \mapsto 
(x^\la,\dot x^0=1,\dot x^i=x^i_0, \dot x^0_\m=K_\m^0,
\dot x^i_\m=K^i_\m).
\ee
It follows that the diagram (\ref{z291}) can be commutative only
if the components $K^0_\m$ of the connection $K$ on $TX\to
X$ vanish. 
Since the coordinate transition functions $x^0\to x'^0$ are independent of
$x^i$, a connection $K$ with the components $K^0_\m=0$ can exist on the tangent
bundle
$TX\to X$. In particular, let $(x^0,x^i)$ be a reference frame. Given
an arbitrary connection $K$ (\ref{cqg3}) on $TX\to X$, one can put 
$K^0_\m=0$ in order to obtain a desired connection
\beq
\ol K= dx^\la\ot (\dr_\la +K^i_\la\dot\dr_i), 
\label{z292}
\eeq
obeying the transformation law
\beq
{K'}_\la^i=(\dr_j x'^i K^j_\m + \dr_\m\dot x'^i)
\frac{\dr x^\m}{\dr x'^\la}.  \label{z293}
\eeq
Now the diagram (\ref{z291}) becomes commutative if the connections
$\g$ and $K$ fulfill the relation
\beq
\g^i_\m= K^i_\m\circ \la=K^i_\m(x^\la,\dot x^0=1, \dot x^i=x^i_0).
\label{z294}
\eeq
It is easily seen that this relation holds globally because the
substitution of $\dot x^i=x^i_0$ into (\ref{z293}) restates the
transformation law (\ref{m175}).  In accordance with the relation
(\ref{z294}), a desired connection $\ol K$ is an extension  of the local
section
$J^1\la\circ \g$ of the affine bundle $J^1_XTX\to TX$ over the closed
submanifold $J^1X\subset TX$ to a global section. Such an extension
always exists, but it is not unique. 
Let us consider the geodesic equation (\ref{cqg11}) on $TX$ with respect to
the connection $\ol K$. Its solution is a geodesic curve $c(t)$ also
satisfying the dynamic equation (\ref{cqg5}), and {\it vice versa}.

\begin{rem}
We can also consider the
injection
$J^2X\to TTX$, given by the coordinate relations
\beq
(x^\la, x^i_0, x^i_{00}) \mapsto (x^\la, \dot x^0 ={\op x^\circ}^0=1,
\dot x^i ={\op x^\circ}^i=x^i_0, \ddot x^0=0, \ddot x^i=x^i_{00}),
\label{cqg80}
\eeq
and show that the dynamic equation (\ref{cqg5}) is equivalent to the 
restriction to (\ref{cqg80}) of a second order equation on $X$. However, one
must prove that this dynamic equation is a geodesic equation. Recall that any
second order dynamic equation 
\beq
\ddot x^\la=\Xi^\la (x^\m,\dot x^\m) \label{cqg110}
\eeq
on $X$ defines a connection 
\beq
K^\la_\m=\frac12\dot \dr_\m\Xi^\la \label{cqg111}
\eeq
on the tangent bundle $TX\to X$ \cite{marmo90,mora}. The equation
(\ref{cqg111}) is a geodesic equation with respect to the connection
(\ref{cqg111}) if and only if it is a spray. It is readily observed that, if
(\ref{cqg110}) is a geodesic equation (\ref{cqg2}) with respect to a
non-linear connection $K$, the corresponding connection (\ref{cqg111}) does
not coincide with $K$ in general.
\end{rem}

\begin{cor}\label{motion1}
In accordance with the relation (\ref{z294}), 
every dynamic equation on the configuration space $X$ can be written in the
form 
\beq
x^i_{00} = K^i_0\circ\la +x^j_0 K^i_j\circ\la, \label{gm340}
\eeq
where $\ol K$ is a connection (\ref{z292}) on the tangent bundle $TX\to X$.
Conversely, 
 each connection $\ol K$ of
the type (\ref{z292}) and, consequently, any connection $K$
(\ref{cqg3}) on the tangent bundle $TX\to X$ defines a connection 
$\g$ on the affine jet bundle $J^1X\to X$ and the dynamic equation
(\ref{gm340}) on the configuration space $X$.
\end{cor}

Proposition \ref{c2} follows at once from the following lemma.

\begin{lem}\label{aff}
There is one-to-one correspondence between the affine connections $\g$ on
the affine jet bundle $J^1X\to X$ and the linear connections $\ol K$
(\ref{z292}) on the tangent bundle $TX\to X$. This correspondence is
given by the relation (\ref{z294}) which takes the form
\ben
&& 
\g^i_\m=\g^i_{\m 0} + \g^i_{\m j}x^j_0 =K_\m{}^i{}_0(x^\la)\dot x^0 + 
K_\m{}^i{}_j(x^\la)\dot x^j|_{\dot x^0=1, \dot x^i=x^i_0}= 
K_\m{}^i{}_0(x^\la) +  K_\m{}^i{}_j(x^\la)x^j_0, \nonumber\\
&& \g^i_{\m\la}= K_\m{}^i{}_\la. \label{gm331}
\een
\end{lem}

In particular, if an affine connection $\g$ is symmetric, so is the
corresponding linear connection $K$.

\begin{cor}\label{aff1}
Any linear connection $K$  on the tangent bundle
$TX\to X$ defines the quadratic dynamic equation
\be
x^i_{00}= K_0{}^i{}_0 +(K_0{}^i{}_j +K_j{}^i{}_0)x^j_0 + K_j{}^i{}_kx^j_0,
x^k_0,
\ee
written with respect to a reference frame $(x^0,x^i)$.
\end{cor}

We conclude this Section with the proof of Proposition \ref{c3}.
Let $\xi$ and $\xi'$ be holonomic connections on the jet bundle $J^1X\to\bR$.
It is readily observed that their difference is a vertical vector field 
\be
\si=(\xi^i-\xi'^i)\dr_i^0
\ee
 which  
takes its values into the vertical tangent bundle $V_XJ^1X\subset VJ^1X$ of the
affine jet bundle $J^1X\to X$.
Similarly to Lemma \ref{aff}, one can show that there is one-to-one
correspondence between the $V_XJ^1X$-valued affine vector fields 
\be
\si=(b^i_k(x^\m)x^k_0 + f^i(x^\m))\dr_i^0
\ee
on $J^1X$ and the $V_XTX$-valued linear vertical vector fields
\be
\ol\si=(b^i_k(x^\m)\dot x^k + f^i(x^\m)\dot x^0)\dot\dr_i
\ee
on TX. This linear vertical field determines a desired soldering form.

\section{Geometry of relativistic mechanics}

Let us consider a mechanical system whose configuration space $X$ has not a
preferable fibration $X\to\bR$. We come to relativistic mechanics on $X$
whose velocity phase space is the  jet manifold of 1-dimensional submanifolds
of $X$; that generalizes the notion of jets of sections of a bundle
\cite{book,mod}.

Let $Z$ be an $(m+n)$-dimensional manifold. 
The 1-order jet manifold $J^1_nZ$ of $n$-dimensional submanifolds of $Z$
comprises the equivalence classes $[S]^1_z$  of
$n$-dimensional imbedded submanifolds of $Z$ which pass through $z\in
Z$,  and are tangent to each other at $z$.
It is provided
with a manifold structure as follows.

Let $Y\to N$ be an $(m+n)$-dimensional bundle over an
$n$-dimensional base $N$ and $\Phi$  an imbedding of $Y$ into $Z$.
Then there is the natural injection
\ben
&& J^1\Phi: J^1Y\to J^1_nZ, \label{cqg60}\\
&& j^1_xs \mapsto [S]^1_{\Phi(s(x))}, \qquad S=\im (\Phi\circ s), \nonumber
\een
where $s$ are sections of $Y\to N$.
This  injection  defines a chart on
$J^1_nZ$.  Such charts with differentiable transition functions cover the set
$J^1_nZ$.
Therefore, one can utilize the following coordinate atlas on the
jet manifold $J^1_nZ$ of submanifolds of $Z$.
Let $Z$ be endowed with a manifold atlas with coordinate charts
\beq
(z^A), \qquad A=1,\ldots,n+m.  \label{5.59}
\eeq
Though $J^0_nZ$, by definition, is diffeomorphic to $Z$, let us provide
$J^0_nZ$ with the atlas where every chart $(z^A)$ on a
domain
$U\subset Z$ is replaced with the
charts on the same domain $U$ which
correspond to the different partitions of the collection $(z^A)$ in
collections of $n$ and $m$ coordinates, denoted by
\beq
(q^\la,y^i), \qquad \la=1,\ldots,n,  \qquad i=1,\ldots,m.\label{5.8}
\eeq
The transition functions between the coordinate charts (\ref{5.8}) of
$J^0_nZ$, associated with the coordinate chart
(\ref{5.59}) of $Z$, reduce simply to exchange between 
coordinates $q^\la$ and $y^i$. 
Transition functions between arbitrary coordinate charts of the
manifold $J^0_nZ$ read
\beq
\wt q^\la = \wt g^\la (q^\mu, y^j), \quad \wt y^i = \wt f^i (q^\mu, y^j),
\quad q^\al=g^\al(\wt q^\m,\wt y^j), \quad y^i=f^i(\wt q^\m,\wt y^j).
\label{5.26}
\eeq

Given the coordinate atlas (\ref{5.8}) of the manifold $J^0_nZ$, the
jet manifold $J^1_nZ$ of $Z$ is endowed with the adapted
coordinates $(q^\la,y^i,y^i_\la)$. 
Using the formal total derivatives $d_\la=\dr_\la +y^i_\la\dr^i$,
one can write the
 transformation rules for these coordinates in the
following form.
 Given the coordinate transformations (\ref{5.26}), it is easy to find that
\beq
d_{\wt q^\la} = \left[d_{\wt q^\la}
g^\al (\wt q^\la, \wt y^i)\right]d_{q^\al}. \label{5.35}
\eeq
Then we have
\beq
\wt y^i_\la =\left[\left(\frac{\dr}{\dr\wt q^\la} + \wt y^p_\la
\frac{\dr}{\dr\wt y^p}\right)
g^\al (\wt q^\la, \wt y^i)\right]\left(\frac{\dr}{\dr q^\al} +y^j_\al
\frac{\dr}{\dr y^j}\right)\wt f^i (q^\mu, y^j). \label{5.36}
\eeq

When $n=1$, the formalism of jets of submanifolds
provides the adequate mathematical description of relativistic mechanics. In
this case, the fibre coordinates $y^i_0$ on $J^1_1Z\to Z$, with the transition
functions (\ref{5.36}), are exactly the familiar
coordinates on  the projective space ${\bf RP}^m$.

Let $X$ be a 4-dimensional world manifold
equipped with an atlas of coordinates $(x^0,x^i)$
(\ref{5.8}) together with the transition functions (\ref{5.26}) which take the
form
\beq
x^0\to \wt x^0(x^0, x^j), \qquad x^i\to \wt x^i(x^0, x^j). \label{b5.1}
\eeq
The coordinates $x^0$ in different charts of $X$ play the role of the temporal
ones. 

Let $J^1_1X$ be the jet manifold of 1-dimensional submanifolds of $X$. It
is provided with the adapted coordinates
$(x^0,x^i,x^i_0)$. Then one can think of $x^i_0$
as being the 3-velocities of relativistic mechanics.  Their
transition functions are obtained as follows.

Given the coordinate transformations (\ref{b5.1}), the total derivative
(\ref{5.35}) reads
\be
d_{\wt x^0}= d_{\wt x^0}(x^0)d_{x^0}=\left(\frac{\dr x^0}{\dr \wt x^0} +
\wt x^k_0\frac{\dr x^0}{\dr \wt x^k}\right)d_{x^0}.
\ee
In accordance with the relation (\ref{5.36}), we have
\be
\wt x^i_0=  d_{\wt x^0}(x^0)d_{x^0}(\wt x^i) =
\left(\frac{\dr x^0}{\dr \wt x^0} +\wt x^k_0\frac{\dr x^0}{\dr \wt
x^k}\right) \left(\frac{\dr \wt x^i}{\dr x^0} +
x^j_0\frac{\dr \wt x^i}{\dr x^j}\right).
\ee
The solution of this equation is
\be
\wt x^i_0= \left(\frac{\dr \wt x^i}{\dr x^0} +
 x^j_0\frac{\dr \wt x^i}{\dr x^j}\right)\mbox{\large /}
 \left(\frac{\dr \wt x^0}{\dr x^0} + x^k_0\frac{\dr \wt x^0}{\dr x^k}\right).
\ee

To obtain the relation between 3- and 4-velocities of a
relativistic system, let us consider the morphism 
\beq
\rho: TX\op\to J^1_1X, \qquad
x^i_0\circ \rho = \dot x^i/\dot x^0. \label{cqg7'}
\eeq
It is readily observed that the coordinate transformation laws of $x^i_0$ and
$\dot x^i/\dot x^0$ are the same. Therefore, one can think of the tangent
bundle
$TX$ as being the space of the 4-velocities of relativistic mechanics.

The morphism (\ref{cqg7'}) is a surjection. Let us assume that the tangent
bundle $TX$ is equipped with a pseudo-Riemannian metric $g$ such that $X$ is
time oriented. The bundle of hyperboloids $W_g$ (\ref{cqg1}) is the disjoint
union 
 of two connected imbedded subbundles of $W^+$ and $W^-$ of $TX$.
Then the restriction
of the morphism (\ref{cqg7'}) to each of these subbundles is an injection
 into $J^1_1X$.

Let us consider the image of this injection in the fibre of $J^1_1X$
over a point $x\in X$. There are coordinates $(x^0,x^i)$ in a neighbourhood
of $x$ such that a pseudo-Riemannian metric $g(x)$ at $x$ is brought 
into the Minkowski one $\eta$.
In this coordinates, the hyperboloid $W_x\subset T_xX$ is 
\be
(\dot x^0)^2 -\op\sum_i (\dot x^i)^2=1.
\ee
This is the union of the subsets $W^+_x$, where $x^0>0$, and $W^-_x$, where
$x^0<0$. The image $\rho(W^+_x)$ is given by the coordinate relation
\be
\op\sum_i (x^i_0)^2<1.
\ee
This relation means that the 3-velocities of a relativistic
system $(X,g)$ are bounded in accordance with the relativity principle.

It should be emphasized the difference between relativistic and
non-relativistic 3-velocities. 
If a world manifold $X$ admits a fibration
$X\to\bR$, there is the canonical imbedding 
\beq
i:J^1X\to J^1_1X \label{cqg31}
\eeq
of the
velocity phase space $J^1X$ of a non-relativistic system on $X\to\bR$ to that
of a relativistic system (cf. (\ref{cqg60})). Moreover, we have
$i=\rho\circ\la$. However, the image of the morphism $i$ (\ref{cqg31}) differs
from that of the morphism (\ref{cqg7'}) for any pseudo-Riemannian metric $g$
on $TX$.

\section{Examples}

In order to compare relativistic and non-relativistic dynamics, one should
consider pseudo-Riemannian metric on $TX$, compatible with the fibration
$X\to\bR$. Note that $\bR$ is a time of non-relativistic mechanics. It
is one for all non-relativistic observers. In the framework of a relativistic
theory, this time can be seen as a cosmological time.  Given  a fibration
$X\to\bR$, a pseudo-Riemannian metric on the tangent bundle $TX$ is said to be
admissible if it is defined by a pair $(g^R,\G)$ of a Riemannian metric on
$X$ and a non-relativistic reference frame $\G$ (\ref{1005}), i.e.,
\ben
&& g=\frac{2\G\ot\G}{\nm\G^2} - g^R, \label{cqg32}\\
&&\nm\G^2=g^R_{\m\nu}\G^\m\G^\nu= g_{\m\nu}\G^\m\G^\nu, \nonumber
\een
in accordance
with the well-known theorem \cite{haw}.
The vector field $\G$ is a time-like vector relative to the
pseudo-Riemannian metric $g$ (\ref{cqg32}), but not with respect to other
admissible pseudo-Riemannian metrics in general.

Given a coordinate systems $(x^0,x^i)$, compatible with the fibration $X\to
\bR$, let us consider a non-degenerate quadratic Lagrangian 
\beq
L=\frac12m_{ij}(x^\m) x^i_0 x^j_0 + k_i(x^\m) x^i_0  +
f(x^\m), \label{cqg20}
\eeq
where $m_{ij}$ is a Riemannian mass tensor. Similarly to
Lemma
\ref{aff}, one can show that any quadratic polynomial in $J^1X\subset TX$ is
extended to a bilinear form in $TX$. Then the Lagrangian $L$
(\ref{cqg20}) can be written as 
\beq
L=-\frac12g_{\al\m}x^\al_0 x^\m_0, \qquad x^0_0=1, \label{cqg40}
\eeq
where $g$ is the metric
\beq
g_{00}=-2f, \qquad g_{0i}=-k_i, \qquad g_{ij}=-m_{ij}. \label{cqg21}
\eeq
The corresponding Lagrange equation takes the form
\beq
x^i_{00}=-(m^{-1})^{ik}\{_{\la k\nu}\}x^\la_0x^\nu_0, \qquad x^0_0=1,
\label{cqg35}
\eeq
where 
\be
\{_{\la\m\nu}\} =-\frac12(\dr_\la g_{\m\nu} +\dr_\nu
g_{\m\la} - \dr_\m g_{\la\nu})
\ee
 are the Christoffel symbols of the metric (\ref{cqg21}). Let us assume that
this metric is non-degenerate. By virtue of
Proposition \ref{c2}, the dynamic equation (\ref{cqg35}) can be brought into
the geodesic equation (\ref{cqg17}) on $TX$ which reads
\ben
&& \dot x^\la\dr_\la \dot x^0 = 0, \qquad \dot x^0=1, \nonumber \\
&&\dot x^\la\dr_\la \dot x^i = \{_\la{}^i{}_\nu\}\dot x^\la\dot x^\nu -
g^{i0}\{_{\la 0\nu}\}\dot x^\la\dot x^\nu. \label{cqg47}
\een

Let us now bring the Lagrangian (\ref{cqg20}) into the form
\beq
L=\frac12m_{ij}(x^\m)( x^i_0-\G^i)(x^j_0-\G^j) +
f'(x^\m), \label{cqg48}
\eeq
where $\G$ is a Lagrangian connection on $X\to\bR$. This connection $\G$
defines an atlas of local constant trivializations of the bundle $X\to\bR$
and the corresponding coordinates $(x^0,\ol x^i)$ on $X$ such that the
transition functions $\ol x^i\to \ol x'^i$ are independent of $x^0$, and
$\G^i=0$ with respect to $(x^0,\ol x^i)$ (see Section 3). In this
coordinates, the Lagrangian $L$ (\ref{cqg48}) reads
\be
L=\frac12\ol m_{ij}\ol x^i_0 \ol x^j_0 +
f'(x^\m). 
\ee
One can think of its first term as the kinetic energy of
a non-relativistic system with the mass tensor $\ol m_{ij}$ relative to the
reference frame
$\G$, while $(-f')$ is a potential.  Let us assume that
$f'$ is a nowhere vanishing function on
$X$. Then the Lagrange equation (\ref{cqg35}) takes the form
\be
\ol x^i_{00}=\{_\la{}^i{}_\nu\}\ol x^\la_0\ol x^\nu_0, \qquad \ol x^0_0=1,
\ee
where $\{_\la{}^i{}_\nu\}$ are the Christoffel symbols of the metric
(\ref{cqg21}) whose components with respect to the coordinates $(x^0,\ol
x^i)$ read
\beq
g_{ij}= -\ol m_{ij}, \qquad g_{0i}=0, \qquad g_{00} =-2f'. \label{cqg51}
\eeq
This metric is Riemannian if $f'>0$ and pseudo-Riemannian if $f'<0$.
 Then the spatial part of the corresponding geodesic
equation
\be
&& \dot {\ol x}^\la\dr_\la \dot{\ol x}^0 = 0, \qquad \dot{\ol x}^0=1,
\\ 
&&\dot {\ol x}^\la\dr_\la \dot {\ol x}^i =
\{_\la{}^i{}_\nu\}\dot{\ol x}^\la\dot{\ol x}^\nu
\ee
is exactly the spatial part of the geodesic equation with
respect to the Levi--Civita connection of the metric (\ref{cqg51}) on $TX$.
It follows that, as was  declared above,  the
non-relativistic dynamic equation (\ref{cqg51}) describes the
non-relativistic approximation of the geodesic motion in the Riemannian or
pseudo-Riemannian space with the metric (\ref{cqg51}). Note that the spatial
part of this metric is the mass tensor which may be treated as a variable
\cite{litt}. 

Conversely, let us consider a geodesic motion
\beq
\dot x^\la\dr_\la \dot x^\m=\{_\la{}^\m{}_\nu\}\dot x^\la\dot x^\nu
\label{cqg70}
\eeq
in the presence of a
pseudo-Riemannian metric $g$ on a world manifold $X$. Let $(x^0,\ol x^i)$ be
local hyperbolic coordinates such that $g_{00}=1$, $g_{0i}=0$. This
coordinates are a non-relativistic frame for a local fibration $X\to\bR$. Then
the equation (\ref{cqg70}) has the non-relativistic limit
\ben
&& \dot{\ol x}^\la\dr_\la \dot{\ol x}^0 = 0, \qquad \dot{\ol x}^0=1, \nonumber
\\ &&\dot{\ol x}^\la\dr_\la \dot{\ol x}^i = \{_\la{}^i{}_\nu\}\dot{\ol
x}^\la\dot{\ol x}^\nu
\label{cqg71}
\een
which is the Lagrange equation for the Lagrangian
\be
L=\frac12\ol m_{ij}\ol x^i_0 \ol x^j_0,
\ee
describing a free 
non-relativistic mechanical system with the mass tensor $\ol m_{ij}=-g_{ij}$.
Relative to another frame $(x^0,x^i(x^0,\ol x^j))$
associated with the same local splitting $X\to\bR$, the non-relativistic limit
of the equation (\ref{cqg70}) keeps the form (\ref{cqg71}), whereas the
non-relativistic equation (\ref{cqg71}) is brought into the
Lagrange equation (\ref{cqg47}) for the Lagrangian
\beq
L=\frac12m_{ij}(x^\m)( x^i_0-\G^i)(x^j_0-\G^j). \label{cqg120}
\eeq
This Lagrangian describes a mechanical system in the presence of the inertial
force associated with the reference frame $\G$. The difference between
(\ref{cqg47}) and (\ref{cqg71}) shows that a gravitational force can not model
an inertial force in general; that depends on both a frame and a system. For
example, if the mass tensor in the Lagrangian $L$ (\ref{cqg120}) is
independent of time, the corresponding Lagrange equation is a spatial part of
the geodesic equation in a pseudo-Riemannian space.

In view of the ambiguity that we have mentioned, the relativization
(\ref{cqg40}) of an arbitrary non-relativistic quadratic Lagrangian
(\ref{cqg20}) may lead to a confusion. In particular, it can be applied to a
gravitational Lagrangian (\ref{cqg48}) where 
\be
f'=-\frac12+\f,
\ee
and $\f$ is a gravitational potential. An arbitrary quadratic dynamic
equation can be written in the form
\be
x^i_{00}=-(m^{-1})^{ik}\{_{\la k\m}\} x^\la_0 x^\m_0 + b^i_\m(x^\nu)x^\m_0,
\qquad x^0_0=1,
\ee
where $\{_{\la k\m}\}$ are the Christoffel symbols of some
admissible pseudo-Riemannian metric $g$, whose spatial part is the mass tensor
$(-m_{ik})$, while
\beq
b^i_k(x^\m) x^k_0 + b^i_0(x^\m) \label{cqg61}
\eeq
is an external force. With respect to the coordinates where  $g_{0i}=0$, one
may construct the relativistic equation
\beq
\dot x^\la\dr_\la\dot x^\m= \{_\la{}^\m{}_\nu\}\dot x^\la\dot x^\nu +
\si^\m_\la \dot x^\la, \label{cqg73}
\eeq
where the soldering form $\si$ must fulfill the condition (\ref{cqg46}). It
takes place only if 
\be
g_{ik}b^i_j + g_{ij}b^i_k=0,
\ee
i.e., the external force (\ref{cqg61}) is the Lorentz-type force plus 
some potential one. Then, we have
\be
\si^0_0= 0, \qquad \si^0_k = -g^{00}g_{kj}b^j_0, \qquad \si^j_k=b^j_k.
\ee

The "relativization" (\ref{cqg73}) exhausts almost all familiar examples.
It means that a wide class of mechanical system can be represented as a
geodesic motion with respect to some affine connection in the spirit of
Cartan's idea. To complete our exposition,  point  out also another
"relativization" procedure. Let a force
$\xi^i(x^\m)$ in the non-relativistic dynamic equation (\ref{cqg5}) be a
spatial part of a 4-vector
$\xi^\la$ in  the Minkowski space $(X,\eta)$. Then one can write the
relativistic equation
\beq
\dot x^\la\dr_\la\dot x^\m= \xi^\la -\eta_{\al\bt}\xi^\bt \dot x^\al\dot
x^\la. \label{cqg74}
\eeq
This is the case, e.g., for a relativistic hydrodynamics that we meet usually
in the literature on a gravitation theory. However, 
the non-relativistic limit $\dot x^0=1$ of (\ref{cqg74}) does not coincide with
the initial non-relativistic equation. There are also other variants of
relativistic hydrodynamic equations \cite{kuper}.

\end{document}